\documentclass[aps,prd,preprint,superscriptaddress]{revtex4-1}
\usepackage{epsfig,dsfont,amssymb,amsmath,amsthm,amsfonts,amsbsy,mathrsfs,mathtools}
\usepackage{graphicx}

\usepackage{float}
\usepackage{color}

\begin{document}

\title{Analytically approximated scalarized black holes and their thermodynamic stability }

\author{Yuan-Xing Gao}
\author{Dao-Jun Liu}
\email{djliu@shnu.edu.cn}
\affiliation{Center for Astrophysics and Department of Physics, Shanghai Normal University, 100 Guilin Road, Shanghai 200234, China}
\begin{abstract}
It is recently shown that, besides the Schwarzshcild black hole solution, there exist also scalarized black hole solutions in some Einstein-scalar-Gauss-Bonnet theories. In this paper, we construct analytical expressions  for the metric functions and scalar field configurations for these scalarized black hole solutions approximately by employing the continued fraction parametrization method and investigate their thermodynamic stability. It is found that the horizon entropy of a scalarized black hole is always smaller than that of a Schwarzschild black hole, which indicates that these scalarized black holes may decay to Schwarzschild black holes by emission of scalar waves. This fact also implies the possibility to extract the energy of scalar charges. 
\end{abstract}
\maketitle

\section{introduction}

In the last few years, the LIGO/Virgo Collaboration has already observed several black hole (BH) mergers and one binary neutron star merger, which implies that we are entering the era of gravitational wave astronomy\cite{Abbott2016,Abbott2017,Collaboration2018}. Even though these observed results are in good agreement with the predictions of General Relativity (GR), one still cannot verify that GR is the preferred theory of gravity. As is well known, there are lots of well-motivated modifications of GR (see e.g. \cite{Clifton2012} for a comprehensive review). However, the majority of their BH solutions are subject to no-hair theorems \cite{Israel1967,Carter1971,Bekenstein1972,Bekenstein1995} that preclude the existence of new charges. For instance, some models contain  scalar degrees of freedom, and the no-hair theorems preclude the existence of BHs with scalar charge. What's interesting is that recent research shows that there is a large class of scalar-tensor theory which can evade the no-hair theorems and admit novel BH solutions with scalar hair ( see \cite{Herdeiro2015} for an early review and \cite{Antoniou2018, PhysRevLett.120.131103, PhysRevLett.120.131104, Bakopoulos2018nui, Doneva:2019vuh, Minamitsuji:2018xde, Minamitsuji:2019iwp, Andreou:2019ikc,   Fernandes:2019rez, Brihaye:2018bgc} for recent developments). The phenomenon, called "spontaneous scalarization", can occur in theories where a scalar field is coupled to the curvature invariant with a $Z_2$-symmetric coupling function and has remarkable features:  BHs in GR are admissible for some constant background scalar field. The coupling with the curvature invariant acts as an effective mass term for the scalar perturbations and can be negative in parts of the BH exterior when the BH mass lies within a certain interval, thus trigger  tachyonic instability and produce a non-zero scalar charge. This tachyonic instability should be quenched by gravitational backreaction in the linear level, or  suppressed by introducing nonlinearity \cite{Silva:2018qhn}, and finally leads to scalarization.

The Einstein-scalar-Gauss-Bonnet (EsGB) theories, as a class  of scalar-tensor theories that possess hairy black hole solutions,   are of very high interest on their own
due to their theoretical motivation coming from attempts to quantize gravity and another two facts:
one is that  GR is not a renormalizable theory but a theory with curvature invariants of second order, such as Gauss-Bonnet invariant, will be renormalizable \cite{Stelle1977}  and the other is that a non-minimal coupling between additional dynamical scalar field and the Gauss-Bonnet invariant would avoid Ostrogradski instability\cite{Berti_2015}. 
Up to now, most of these scalarized BH solutions are obtained numerically in varies of EsGB theories. Especially, it is shown that when the scalar field is coupled to Gauss-Bonnet invariant with a quadratic coupling or an exponential coupling, the static scalarized solutions appears.  However, it is not convenient to analyze the properties of these numerical solutions  in detail, although some arguments on the stability of these BHs have been already pointed out \cite{Antoniou2018}.   Needless to say, if  analytical expressions of these scalarized black hole solutions are obtained, it would be much more convenient to investigate the properties of these black hole solutions.    In the present paper, we shall revisit  the thermodynamic stability of the spherically-symmetric scalarzied black holes in the quadratic case of the EsGB theory by  constructing analytical expressions for the numerical solutions via the parametrization method first proposed by Rezzolla and  Zhidenko \cite{Rezzolla:2014mua}  which uses a continued-fraction expansion in terms of a compactified radial coordinate. In fact, it has been shown that this method can obtain  analytical expressions with adequate precision for such numerical black hole metrics, which are valid  in the whole space outside the event horizon in various circumstances \cite{Kokkotas:2017ymc,Kokkotas2017a,Konoplya:2019goy,Konoplya:2019fpy}. 

The paper is organized as follows. In Sec.\ref{sec:2}, a brief review on the EsGB theory of gravity with a quadratic coupling and the regime of spontaneous scalarization are given. Next, the numerical method employed is described and the numerical results are presented in Sec.\ref{sec:3}. Then, the parameterized method applied to the numerical solutions is performed in Sec.\ref{sec:4}, and thermodynamics quantities for scalarized black holes are shown in Sec.\ref{sec:5}. Finally, we conclude in Sec.\ref{sec:conclusion}. Throughout this paper, we use geometric units with $c = G = 1$ and the mostly plus metric signature.

\section{Spontaneous scalarization}
\label{sec:2}
Let us consider a sort of EsGB theories of which the action is given by
\begin{equation}\label{eq:action}
\begin{aligned}
S_{EsGB}=\frac{1}{16\pi}\int d^4x\sqrt{-g}\left[R-\frac{1}{2}\nabla_\mu\phi\nabla^\mu\phi+\alpha\phi^2R_{GB}^2 \right]
\end{aligned}
\end{equation}
where $\alpha$ is the coupling constant and $R_{GB}^2$  the Gauss-Bonnet invariant
\begin{equation}
R_{GB}^2=R^2-4R_{\mu\nu}R^{\mu\nu}+R_{\mu\nu\rho\sigma}R^{\mu\nu\rho\sigma}.
\end{equation} 
Varying the action with respect to $g_{\mu\nu}$ and $\phi$ lead to the dynamical equations for the metric  
\begin{equation}\label{eq:Einstein}
G_{\mu\nu}=\frac{1}{2}\partial_\mu\phi\partial_\nu\phi-\frac{1}{4}g_{\mu\nu}(\partial\phi)^2-4\alpha\nabla^\rho\nabla^\sigma(\phi^2)P_{\mu\rho\nu\sigma}
\end{equation}
and the scalar field
\begin{equation}\label{eq:C-tensor}
(\Box+2\alpha R_{GB}^2)\phi=0,
\end{equation}
respectively. Here the tensor $P_{\mu\nu\rho\sigma}$ takes the form
\begin{equation}
P_{\mu\rho\nu\sigma}=R_{\mu\rho\nu\sigma}+g_{\mu\sigma}R_{\nu\rho}-g_{\mu\nu}R_{\rho\sigma}+g_{\nu\rho}R_{\mu\sigma}-g_{\rho\sigma}R_{\mu\nu}+\frac{R}{2}(g_{\mu\nu}g_{\rho\sigma}-g_{\mu\sigma}g_{\nu\rho}).
\end{equation}
Note that if we choose a background quantity $\bar{\phi}=0$, Eqs.(3) and (4) admit the Schwarzschild solution
\begin{equation}
ds^2=-f(r)dt^2+\frac{dr^2}{f(r)}+r^2(d\theta^2+\sin^2\theta d\varphi^2)
\end{equation}
with $f(r)=1-\frac{2M}{r}$. In this case, if perturbations around the background are introduced as
\begin{equation}
g_{\mu\nu}=\bar{g}_{\mu\nu}+h_{\mu\nu},\ \ \phi=\bar{\phi}+\delta\phi,
\end{equation}
then the metric perturbation $h_{\mu\nu}$ will decay quickly and finally vanish in the background of Schwarzschild spacetime, since it is decoupled with the scalar field on the linear level. The remaining linearized equation is just for the perturbed scalar field $\delta\phi$
\begin{equation}
\left(\bar{\Box}+\frac{96\alpha M^2}{r^6}\right)\delta\phi=0.
\end{equation}
Decomposing $\delta\phi$ with spherical harmonics $Y_{lm}$ as 
\begin{equation}
\delta\phi=e^{-i\omega t}\frac{u(r)}{r}Y_{lm},
\end{equation} 
the radial equation is given by 
\begin{equation}
\frac{d^2u}{dr_*^2}+[\omega^2-V(r)]u=0,
\end{equation}
with $V(r)=f(r)\left[\frac{l(l+1)}{r^2}+\frac{2M}{r^3}-\frac{96\alpha M^2}{r^6} \right]$ and $dr_*=dr/f(r)$ is the tortoise radial coordinate. 
\par As shown by many authors, $u(r)$ becomes unstable in certain regions (for instance, it is found that $\alpha>r_+^2/11.03$ in  \cite{Myung:2019wvb} ), which indicates that the perturbed scalar field will grow continuously and then take a backreaction on the spacetime geometry. Consequently, novel black holes with non-trivial scalar configuration are produced in this regime. However, it is hard to derive analytical representations for these scalarized black holes directly, so we shall obtain these solutions numerically and construct an analytical approximation for them by using the parameterization method proposed in Ref.\cite{Rezzolla:2014mua}.

\section{The numerical  scalarized black hole solution}\label{sec:3}
In the spontaneous scalarization context, we seek static spherically symmetric solutions with line element 
\begin{equation}\label{eq:metric11}
ds^2=-A(r)dt^2+\frac{dr^2}{B(r)}+r^2(d\theta^2+\sin^2\theta d\varphi^2),
\end{equation}   
where the pending functions $A(r)$ and $B(r)$ are independent. 
By employing the line element \eqref{eq:metric11}, the ($t$, $t$) and($r$, $r$) components of Einstein's equations take the explicit form as
\begin{equation}\label{eq:metricA}
2B'[r+4\alpha(1-3B)\phi\phi']+\frac{1}{2}[B(r^2\phi'^2+4)-4]-16\alpha B(B-1)(\phi'^2+\phi\phi'')=0,
\end{equation}
\begin{equation}\label{eq:metricB}
2A'[r+4\alpha(1-3B)\phi\phi']-\frac{A}{2B}[B(r^2\phi'^2-4)+4]=0.
\end{equation}
While the  equation for the scalar field reads
\begin{equation}
\phi''+\left(\frac{2}{r}+\frac{A'}{2A}+\frac{B'}{2B}\right)\phi'=\frac{4\alpha\phi A'}{r^2A^2B}(A'B^2+AB'-A'B-3ABB')-\frac{8\alpha\phi}{r^2A}(B-1)A''.
\end{equation}
Equation \eqref{eq:metricB} can be treated as a second-order polynomial with respect to $B$, which is solved algebraically: 
\begin{equation}\label{eq:phi}
\frac{1}{B}=\frac{4A'(r+4\alpha\phi\phi')+A(4-r^2\phi'^2)+\sqrt{[4A'(r+4\alpha\phi\phi')+A(4-r^2\phi'^2)]^2-768\alpha AA'\phi\phi'}}{8A}.
\end{equation}
Here, the (-) branch has already been excluded since it doesn't contribute to the formation of scalarized black holes. Then, eliminating $B$ from Eqs.\eqref{eq:metricA} and \eqref{eq:phi}, one can form a system of two independent ordinary differential equations of second order for  $A$ and $\phi$:
\begin{equation}\label{eq:16}
\begin{aligned}
A''&=F_A(A,A',\phi,\phi'; r),\\
\phi''&=F_\phi(A,A',\phi,\phi'; r).
\end{aligned}
\end{equation}
Here the complicated expressions for $F_A$ and $F_\phi$ are not written explicitly. Once we derive the solution for $A$ and $\phi$, the solution for $B$ can be obtained immediately via Eq. \eqref{eq:phi}. 

%\subsection{Boundary conditions}
%\subsubsection{Near the horizon} 
Near the horizon, let us consider power-law expansion of the solution in terms of $(r-r_0)$ as
\begin{equation}\label{eq:17}
\begin{aligned}
A&=\sum^{\infty}_{n=1}a_n(r-r_0)^n,\\
\phi&=\sum^{\infty}_{n=0}\phi_n(r-r_0)^n.
\end{aligned}
\end{equation}
Substituting \eqref{eq:17} into \eqref{eq:16}, one can find that equations for $\{a_1, \phi_0, \phi_1\}$ obtained order by order vanish which satisfy the requirements in the vicinity of the horizon automatically, while the set for $\{a_2, \phi_2\}$ is given by
\begin{eqnarray}\label{eq:18-1}
&&3r_0^3a_1\phi_1^2+r_0^2(12\alpha a_1\phi_0\phi_1^3-8a_2)+16\alpha\phi_0\phi_1[a_1-8\alpha a_2\phi_0\phi_1+4\alpha a_1(\phi_1^2+2\phi_0\phi_2)]\nonumber\\
&&+8r_0[a_1(2\alpha(\phi_1^2+2\phi_0\phi_2)-1)-8\alpha a_2\phi_0\phi_1]=0,
\end{eqnarray}
and
\begin{eqnarray}\label{eq:18-2}
&&r_0^4a_1\phi_1+7r_0^3\alpha a_1\phi_0\phi_1^2+4r_0^2\alpha\phi_0(3\alpha a_1\phi_0\phi_1^3-2a_2)+16r_0\alpha\phi_0[a_1-4\alpha a_2\phi_0\phi_1+\alpha a_1(\phi_1^2+2\phi_0\phi_2)]\nonumber\\
&&+16\alpha^2\phi_0^2\phi_1[a_1-8\alpha a_2\phi_0\phi_1+4\alpha a_1(\phi_1^2+2\phi_0\phi_2)]=0.
\end{eqnarray}
Note that Eqs. \eqref{eq:18-1} and \eqref{eq:18-2} are not self-consistent unless the following condition is satisfied:
\begin{equation}
4r_0^2\alpha \phi_0\phi_1^2+r_0^3\phi_1+24\alpha\phi_0=0,
\end{equation}
which can be easily solved to yield
\begin{equation}
\phi_1=\frac{-r_0^2\pm\sqrt{r_0^4-384\alpha^2\phi_0^2}}{8\alpha\phi_0r_0}
\end{equation}
and should be treated as a constraint for $\phi_1$. We will choose the (+) branch, since it covers $\phi'(r_0)=0$ which corresponds to the requirement of the no-hair theorem which implys that a vanishing coupling leads to a trivial scalar field. Besides, to ensure $\phi_1$ is real, the following constraint should be imposed for $\phi_0$ \begin{equation}
\phi_0^2<\frac{r_0^4}{384\alpha^2}.
\end{equation}
For  the fixed value of $\alpha$ and $r_0$, one can construct  black hole solutions for each value of $\phi_0$ in the range $\left(-\frac{r_0^2}{8\sqrt{6}\alpha},\ \frac{r_0^2}{8\sqrt{6}\alpha}\right)$. In the rest of the paper, we only consider $\phi_0$ of which the value falls into the range $\left(0,\ \frac{r_0^2}{8\sqrt{6}\alpha}\right)$, since another half range will lead to similar results.
Consequently, we obtained the asymptotic solution for $\{A, B, \phi\}$ by the expressions
\begin{equation}
\begin{aligned}
A(r)&=a_1(r-r_0)+\mathcal{O}((r-r_0)^2),\\
B(r)&=\frac{r_0(r_0^2-\sqrt{r_0^4-384\alpha^2\phi_0^2})}{192\alpha^2\phi_0^2}(r-r_0)+\mathcal{O}((r-r_0)^2),\\
\phi(r)&=\phi_0+\frac{-r_0^2+\sqrt{r_0^4-384\alpha^2\phi_0^2}}{8\alpha\phi_0r_0}(r-r_0)+\mathcal{O}((r-r_0)^2),
\end{aligned}
\end{equation}
where $\phi_0$ is the amplitude of the scalar field at the horizon. Note that $a_1$ is an arbitrary constant, of which the value is fixed by matching the asymptotic of $A$ at infinity.  

%\subsubsection{At spatial infinity}
At spatial infinity, the metric functions and the scalar field can be again expanded in terms of $1/r$. Associated with the ADM mass $M$ and scalar charge $D$, the following expressions can be derived:
\begin{equation}
\begin{aligned}
A&=1-\frac{2M}{r}+\frac{MD^2}{12r^3}+\mathcal{O}(\frac{1}{r^4}),\\
B&=1-\frac{2M}{r}+\frac{D^2}{4r^2}+\frac{MD^2}{4r^3}+\mathcal{O}(\frac{1}{r^4}),\\
\phi&=\phi_\infty+\frac{D}{r}+\mathcal{O}(\frac{1}{r^2}).
\end{aligned}
\end{equation}
%\subsection{Mass and scalar charge}
From (23), the mass $M$ and the scalar charge $D$ can be evaluated as
\begin{equation}
\begin{aligned}
M&=\frac{r}{2}(1-B)|_{r\to\infty},\\
D&=-r^2\phi'|_{r\to\infty},
\end{aligned}
\end{equation}
which are useful for later calculation.

\begin{figure}        
	\centering 
	\includegraphics[width=0.7\textwidth]  {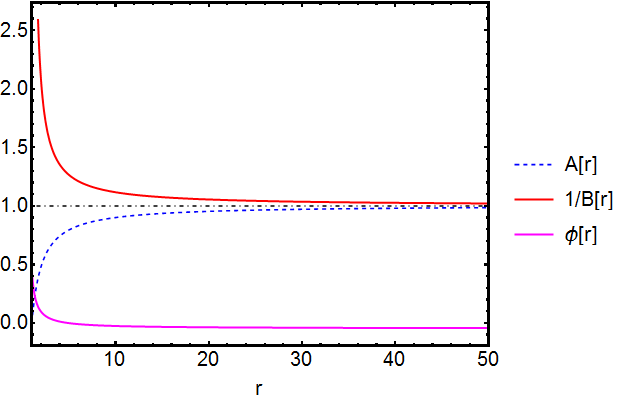}          
	\caption{$A(r), 1/B(r)$ and $\phi(r)$ as functions of $r$ for $\alpha=0.1, r_0=1, \phi_0=0.5$.}
	\label{fig:1}
\end{figure}

%\subsection{Scalarized black holes}
To simplify the numerical calculation, we fix $\alpha=0.1$ and $r_0=1$, thus the radial coordinate is measured in the units of the horizon radius. Then by employing a standard shooting method, we obtained the numerical solution for $\phi_0=0.5$, which is shown in Fig.\ref{fig:1}. Note that the asymptotic value of $\phi$ at infinity is about $-0.047$ rather than $0$.

\section{Analytical approximation}
\label{sec:4}
\subsection{Construction of the analytical formula}
Following \cite{Rezzolla:2014mua, Kokkotas:2017ymc}, we introduce a compactified coordinate $x$:
\begin{equation}
x=1-\frac{r_0}{r},
\end{equation}
so that $x = 0$ corresponds to the event horizon and $x = 1$ to infinity. With this new coordinate, Eqs.(15) and (16) can be rewritten in terms of $x$ as:
\begin{equation}\label{eq:26}
B(x)=\frac{G(x)-\sqrt{G(x)^2-768\alpha r_0^2(1-x)^4A(x)A'(x)\phi(x)\phi'(x)}}{96\alpha(1-x)^4A'(x)\phi(x)\phi'(x)}
\end{equation}
and
\begin{equation}\label{eq:27}
\begin{aligned}
A''(x)&=F_A(A,A',\phi,\phi';x),\\
\phi''(x)&=F_\phi(A,A',\phi,\phi';x),
\end{aligned}
\end{equation}
where $G(x)=4A'(x)[r_0(1-x)+4\alpha(1-x)^4\phi(x)\phi'(x)]+r_0^2A(x)[4-(1-x)^2\phi'(x)^2]$.\\
Subsequently, we define there new functions $J(x)$, $K(x)$ through these relations:
\begin{equation}
A(x)=xJ(x),
\end{equation}
\begin{equation}
\sqrt{\frac{A(x)}{B(x)}}=K(x).
\end{equation}
The representations of the above functions are taken as follows:
\begin{equation}
\begin{aligned}
J(x)&=1-\epsilon(1-x)+(j_0-\epsilon)(1-x)^2+\tilde{J}(x)(1-x)^3,\\
K(x)&=1+k_0(1-x)+\tilde{K}(x)(1-x)^2,\\
\end{aligned}
\end{equation}
and similarly 
\begin{equation}
\phi(r(x))=\phi_\infty+l_0(1-x)+\tilde{L}(x)(1-x)^2,
\end{equation}
where $\tilde{J}(x)$, $\tilde{K}(x)$ and  $\tilde{L}(x)$ are given in terms of the continued fractions to describe the metric near the horizon:
\begin{equation}
\begin{aligned}
\tilde{J}(x)&=\frac{j_1}{1+\frac{j_2x}{1+\frac{j_3x}{1+\frac{j_4x}{1+...}}}},\\
\tilde{K}(x)&=\frac{k_1}{1+\frac{k_2x}{1+\frac{k_3x}{1+\frac{k_4x}{1+...}}}},\\
\tilde{L}(x)&=\frac{l_1}{1+\frac{l_2x}{1+\frac{l_3x}{1+\frac{l_4x}{1+...}}}}.
\end{aligned}
\end{equation}

\begin{figure}        
	\centering 
	\includegraphics[width=0.45\textwidth]  {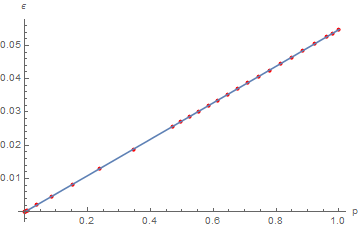}  
	\includegraphics[width=0.45\textwidth]  {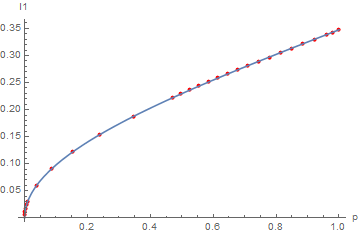}     
	\caption{Fit (solid line) of $\epsilon$ (left panel) and $l_1$ (right panel) as functions of $p$.} \label{fig:1a}  
\end{figure}

The coefficients $\epsilon, j_0, k_0$ and $l_0$ are introduced to match the asymptotic behavior at infinity (23). Thus,
\begin{equation}
%\begin{aligned}
\epsilon=-\left(1-\frac{2M}{r_0}\right),\quad
j_0=0 ,\quad
k_0=0 ,\quad
l_0=\frac{D}{r_0}.
%\end{aligned}
\end{equation}
Then, we truncate the third coefficients $j_3=k_3=l_3=0$, since in the spontaneous scalarization context, a second order approximation is sufficient. From Eqs.\eqref{eq:26} and \eqref{eq:27}, one can express the coefficients $j_2, k_2$ as functions of $j_1,l_1, l_2$, of which the values can be derived by comparisons with the asymptotic at event horizon \eqref{eq:17}:
\begin{equation}
\begin{aligned}
&j_1=r_0a_1+2\epsilon-1,\\
&l_1=\frac{r_0(\phi_0-\phi_\infty) -D}{r_0},\\
&l_2=-\frac{r_0^2\phi_1+D}{r_0(\phi_0-\phi_\infty)-D}-2 .
\end{aligned}
\end{equation}
Here it is advantageous to define a dimensionless parameter $p$ via
\begin{equation}
p=\frac{384\alpha^2\phi_0^2}{r_0^4},\ \ \ \ 0<p<1.
\end{equation}
At the second order of expansion, the coefficients $\epsilon, j_1, j_2, k_1, k_2, l_0, l_1, l_2$ are best fit by the functions of $p$ as follows
\begin{eqnarray}
\epsilon &=& -0.0000564151 + 0.0544789 p + 0.000368755 p^2 ,\\
j_1&=& 0.0060185 + 0.0260605 p - 7.07155 p^2 + 98.8815 p^3 - 697.603 p^4 + 
2803.02 p^5 - 6825.37 p^6\nonumber\\
& &+ 10256.5 p^7 - 9295.29 p^8 + 
4657.01 p^9 - 990.476 p^{10} , \\
j_2&=& -55470.8 p + 739468 p^2 - 4.34496\times10^6 p^3 + 1.4772\times10^7 p^4 - 
3.20274\times10^7 p^5 \nonumber\\
&&+ 4.59271\times10^7 p^6 - 4.35644\times10^7 p^7 + 
2.63618\times10^7 p^8 - 9.23567\times10^6 p^9\nonumber\\
&& + 1.42753\times10^6 p^{10}, \\
k_1&=& 0.00299087 - 0.124225 p - 7.36854 p^2 + 149.869 p^3 - 1563 p^4 + 
9530.87 p^5 - 36651.8 p^6 \nonumber\\
&&+ 92476.7 p^7 - 155302 p^8 + 
171983 p^9 - 120626 p^{10} + 48551.2 p^{11} - 8541.9 p^{12} ,\\
k_2&=& -106881 p + 1.42403\times10^6 p^2 - 8.36139\times10^6 p^3 + 2.84058\times10^7 p^4 - 
6.15398\times10^7 p^5  \nonumber\\
&& +8.81791\times10^7 p^6- 8.35779\times10^7 p^7 + 
5.0536\times10^7 p^8 - 1.76915\times10^7 p^9 \nonumber\\
&&+ 2.73248\times10^6 p^{10} , 
\end{eqnarray}
\begin{eqnarray}
l_0&=&0.237955 \sqrt{p} + 0.00010036 p - 0.026873 p^2 , \\
l_1&=&0.30475 \sqrt{p} + 0.0140045 p + 0.028906 p^2 , \\
l_2&=&1.19334 + 1.18938 \sqrt{p} - 13.6844 p + 490.427 p^2 - 10347.3 p^3 + 
 126174. p^4 - 959388. p^5 \nonumber\\
 &&+ 4.82418\times10^6 p^6 - 1.66709\times10^7 p^7 + 
 4.04867\times10^7 p^8 - 6.96964\times10^7 p^9\nonumber\\
  && + 8.46005\times10^7 p^{10}-7.08019\times10^7 p^{11} + 3.8878\times10^7 p^{12}\nonumber\\
  && - 1.26064\times10^7 p^{13} + 1.82939\times10^6 p^{14}.
\end{eqnarray}
%\end{equation}
%For each $p$, we calculate numerically the values needed in Eqs.(35)-(37) to obtain the coefficients $\epsilon, l_0, j_1, l_1, l_2$, and derive further $j_2, k_2$.
%\newpage
The accuracy for $\epsilon$ and $l_1$ can be seen in Fig. \ref{fig:1a}.
Since the value of $\phi_\infty$ is uniquely determined by each $p$, we also fit $\phi_\infty$ by
\begin{eqnarray}
\phi_\infty&=& -0.0354014 \sqrt{p} - 0.00407199 p - 0.0139096 p^2 - 0.0171015 p^3 \nonumber\\
&&+  0.0713693 p^4  - 0.0855375 p^5  + 0.0365507 p^6.
\end{eqnarray}
Finally, the analytical approximate expressions $A(r)_p, B(r)_p$ for  the metric functions $A(r), B(r)$, with all the coefficients above, get the following form  
\begin{equation}
\label{eq:approx1}
\begin{split}
A(r)_p=&\left(1-\frac{r_0}{r}\right)\left(1 - \frac{0.000368755 (-0.152988 + 147.738 p + p^2) r_0}{r}\right.\\
& \left.+ \frac{(0.0000564151 - 0.0544789 p - 0.000368755 p^2) r_0^2}{r^2}+\frac{Q_1}{T_1}\right) ,
\end{split}
\end{equation}
\begin{equation}\label{eq:approx1.2}
B(r)_p=\frac{Q_2}{T_2} ,
\end{equation}
where
\begin{eqnarray}
Q_1&=& -0.000693838 r_0^3(-6.07637\times10^{-6} - 0.0000263111 p + 0.00713955 p^2 - 0.0998323 p^3\nonumber\\
&& + 0.704311 p^4 - 2.82998 p^5 + 6.891 p^6 - 10.3551 p^7 + 9.38466 p^8 - 4.70179 p^9 + p^{10}) ,
\end{eqnarray}
\begin{eqnarray}
T_1&=& r^2 ((7.0051\times10^{-7} - 0.0388578 p + 0.518004 p^2 - 3.04369 p^3 + 10.348 p^4 - 22.4355 p^5 + 32.1723 p^6\nonumber \\
&&- 30.5173 p^7 + 18.4667 p^8 - 6.46967 p^9 + p^{10}) r + p(0.0388578 - 0.518004 p + 3.04369 p^2\nonumber \\
&&- 10.348 p^3 + 22.4355 p^4 - 32.1723 p^5 + 30.5173 p^6 - 18.4667 p^7 + 6.46967 p^8 - p^9) r_0),
\end{eqnarray}
\begin{eqnarray}
Q_2&=&(r-r_0) ((3.65968\times10^{-7} - 0.0391149 p + 0.521148 p^2 - 3.06 p^3 + 10.3956 p^4 - 22.5216 p^5 \nonumber\\
&&+ 32.2707 p^6 - 30.5868 p^7 + 18.4945 p^8 - 6.47451 p^9 + p^{10}) r + p (0.0391149 - 0.521148 p + 3.06 p^2  \nonumber\\
&&-10.3956 p^3 + 22.5216 p^4 - 32.2707 p^5 + 30.5868 p^6 - 18.4945 p^7 + 6.47451 p^8 - p^9) r_0)^2  \nonumber\\
&&((7.0051\times10^{-7} - 0.0388578 p + 0.518004 p^2 - 3.04369 p^3 + 10.348 p^4 - 22.4355 p^5 + 32.1723 p^6  \nonumber\\
&&- 30.5173 p^7 + 18.4667 p^8 - 6.46967 p^9 + p^{10}) r^3 + (3.95193\times10^{-11} + 0.0388556 p - 0.515858 p^2  \nonumber\\
&&+ 3.01531 p^3 - 10.1818 p^4 + 21.8716 p^5 - 30.9521 p^6 + 28.7711 p^7 - 16.815 p^8 + 5.47452 p^9  \nonumber\\
&&- 0.654292 p^{10} - 0.0520932 p^{11} - 0.000368755 p^{12}) r^2 r_0 + (3.95193\times10^{-11} - 3.8163\times10^{-8} p \nonumber\\
&&- 2.58316\times10^{-10} p^2) r r_0^2 + (4.21602\times10^{-9} + 2.21042\times10^{-6} p - 0.00215111 p^2 + 0.028447 p^3  \nonumber\\
&&- 0.166698 p^4 + 0.565853 p^5 - 1.22504 p^6 + 1.75335 p^7 - 1.65824 p^8 + 0.99842 p^9 - 0.346402 p^{10} \nonumber\\
&&+ 0.0520932 p^{11} + 0.000368755 p^{12}) r_0^3) ,
\end{eqnarray}
\begin{eqnarray}
T_2&=& r ((7.0051\times10^{-7} - 0.0388578 p + 0.518004 p^2 - 3.04369 p^3 + 10.348 p^4 - 22.4355 p^5 + 32.1723 p^6 \nonumber\\
&&- 30.5173 p^7 + 18.4667 p^8 - 6.46967 p^9 + p^{10}) r + p (0.0388578 - 0.518004 p + 3.04369 p^2 \nonumber\\
&&- 10.348 p^3 + 22.4355 p^4 - 32.1723 p^5 + 30.5173 p^6 - 18.4667 p^7 + 6.46967 p^8 - p^9) r_0) \nonumber\\
&&((3.65968\times10^{-7} - 0.0391149 p + 0.521148 p^2 - 3.06 p^3 + 10.3956 p^4 - 22.5216 p^5 + 32.2707 p^6 \nonumber\\
&&- 30.5868 p^7 + 18.4945 p^8 - 6.47451 p^9 + p^{10}) r^2 + p(0.0391149 - 0.521148 p + 3.06 p^2 \nonumber\\
&&- 10.3956 p^3 + 22.5216 p^4 - 32.2707 p^5 + 30.5868 p^6 - 18.4945 p^7 + 6.47451 p^8 - p^9) r r_0 \nonumber\\
&&+ (1.09456\times10^{-9} - 4.54624\times10^{-8} p - 2.69665\times10^{-6} p^2 + 0.0000548474 p^3 - 0.000572006 p^4 \nonumber\\
&&+ 0.00348799 p^5 - 0.0134134 p^6 + 0.0338435 p^7 - 0.0568356 p^8 + 0.0629402 p^9 - 0.0441451 p^{10} \nonumber\\
&&+ 0.0177682 p^{11} - 0.00312606 p^{12}) r_0^2)^2 .
\end{eqnarray}
Meanwhile, the analytical approximate expression $\phi(r)_p$ for the scalar field $\phi(r)$ reads
\begin{eqnarray}
\label{eq:approx2}
\phi(r)_p&=&-0.0354014 \sqrt{p} - 0.00407199 p - 0.0139096 p^2 - 0.0171015 p^3 + 0.0713693 p^4 - 0.0855375 p^5 \nonumber\\
&&+ 0.0365507 p^6 + \frac{(0.237955 \sqrt{p} + 0.00010036 p - 0.026873 p^2) r_0}{r}+\frac{Q_3}{T_3}
\end{eqnarray}
where
\begin{equation}
Q_3=(0.30475 \sqrt{p} + 0.0140045 p + 0.028906 p^2) r_0^3 ,
\end{equation}
\begin{eqnarray}
T_3&=&r^3 + (1.19334 + 1.18938 \sqrt{p} - 13.6844 p + 490.427 p^2 - 10347.3 p^3 + 126174p^4 \nonumber\\ 
&&- 959388p^5 
+ 4.82418\times10^6 p^6 -1.66709\times10^7 p^7 + 4.04867\times10^7 p^8 - 6.96964\times10^7 p^9 \nonumber\\
&& + 8.46005\times10^7 p^{10} - 7.08019\times10^7 p^{11} + 3.8878\times10^7 p^{12} - 1.26064\times10^7 p^{13} \nonumber\\
&& + 1.82939\times10^6 p^{14}) (r^3 -r_0r^2).
\end{eqnarray}

\subsection{Error analysis}
\begin{figure}      
	\centering 
	\includegraphics[width=0.7\textwidth]  {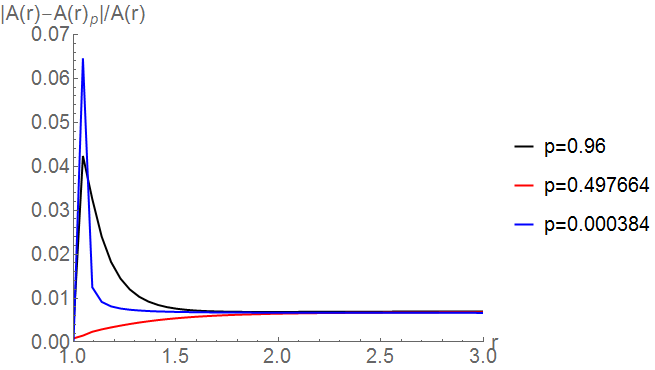}          
	\caption{The relative deviation between the numerical value of $A(r)$ and its analytical approximation $A(r)_p$ for $\alpha=0.1, r_0=1$.}\label{fig:3}
\end{figure}

\begin{figure}        
	\centering 
	\includegraphics[width=0.7\textwidth]  {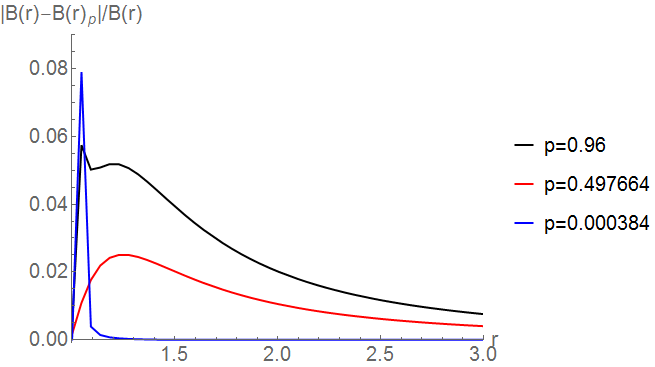}          
	\caption{The relative  deviation between the numerical value of $B(r)$ and its analytical approximation $B(r)_p$ for $\alpha=0.1, r_0=1$.}\label{fig:4}
\end{figure}

\begin{figure}        
	\centering 
	\includegraphics[width=0.7\textwidth]  {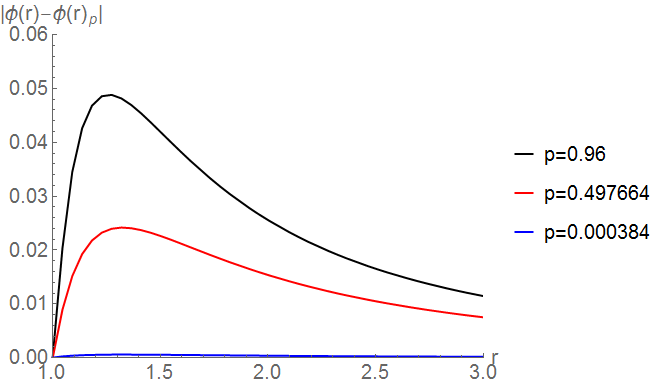}          
	\caption{The  deviation between the numerical value of $\phi(r)$ and its analytical approximation $\phi(r)_p$ for $\alpha=0.1, r_0=1$.}\label{fig:5}
\end{figure}

The errors between our analytic approximate representations \eqref{eq:approx1}, \eqref{eq:approx1.2} and \eqref{eq:approx2} and the numerical solution $A(r)$, $B(r)$ and $\phi(r)$ are shown in Fig. \ref{fig:3}, \ref{fig:4} and \ref{fig:5}, respectively. As mentioned above, we choose $\alpha=0.1$, $r_0=1$. For the metric function $A(r)$ and $B(r)$, in the whole region outside the event horizon, the analytical representations fit well with the numerical solution with relative error of a fraction lower than 2 percent, and it is worth mentioning that in the region $r>3$, the relative error is about 0.7 percent. However, it is clear that the relative error increases near the horizon when $p$ approaches its extremal values $0$ and $1$. For the scalar field $\phi(r)$, we calculate the absolute error rather than relative error, since there is a zero near the horizon.

\section{thermodynamic quantities from the analytical approximation}
\label{sec:5}
With the approximate analytical representations \eqref{eq:approx1}, \eqref{eq:approx1.2} and \eqref{eq:approx2} in hand, one can easily compute some interesting physical quantities. 
\subsection{Hawking temperature}
\begin{figure}        
	\centering 
	\includegraphics[width=0.6\textwidth]  {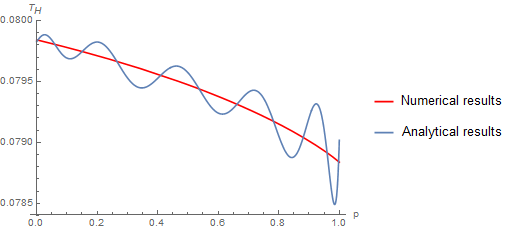}          
	\caption{The Hawking temperature $T_H$ as a function of $p$.}\label{fig:6}
\end{figure}

In the scalarized black hole spacetime, the Killing vector and static four-velocity are defined by
\begin{equation}
K^\mu=(1, 0, 0, 0),\ \ U^\mu=\left(\frac{1}{\sqrt{A(r)}}, 0, 0, 0 \right),
\end{equation}
so the redshift factor is 
\begin{equation}
V=\sqrt{-K_\mu K^\mu}=\sqrt{A(r)}.
\end{equation}
Using the relation $a_\mu=\nabla_\mu\mathrm{ln}V$, the acceleration reads 
\begin{equation}
a_\mu=\frac{A'(r)}{2A(r)}\nabla_\mu r,
\end{equation}
with the magnitude 
\begin{equation}
a=\frac{A'(r)}{2A(r)}\sqrt{B(r)}. 
\end{equation}
Thus, the surface gravity and the Hawking temperature of the scalarized black holes are 
\begin{equation}
\kappa_H=Va=\frac{A'(r)}{2}\sqrt{\frac{B(r)}{A(r)}},
\end{equation}
\begin{equation}\label{eq:TH}
T_H=\frac{\kappa_H}{2\pi}=\frac{A'(r)}{4\pi}\sqrt{\frac{B(r)}{A(r)}},
\end{equation}
respectively.
Plugging \eqref{eq:approx1} and \eqref{eq:approx1.2} into \eqref{eq:TH}, one can derive the representation for $T_H$ as functions of $p$. For each $p$, the value of $T_H$ is shown in Fig.\ref{fig:6}.
Compared to numerical results, though oscillating behaviors are included in the results of our analytical expressions, the relative error is quite small, within 0.5 percent.

\subsection{The horizon entropy}
\begin{figure}        
	\centering 
	\includegraphics[width=0.5\textwidth]  {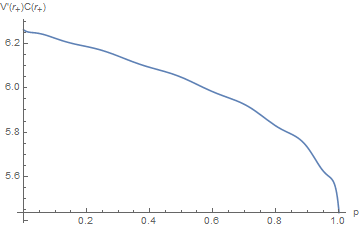}          
	%\caption{The case $p=0$ corresponds to the Schwarzschild solution, the integrand $V'(r_+)C(r_+)$ decreases as $p$ increases.}
	\caption{The integrand $V'(r_+)C(r_+)$ as a function of $p$ and the case $p=0$ corresponds to the Schwarzschild solution.}
	\label{fig:7}
\end{figure}

When assuming a horizon equation of state, one can get a horizon first law by considering a virtual displacement, from which the entropy can be obtained \cite{Padmanabhan2002}. Adding the contribution of matter and assuming that the sources of thermodynamic system are also the sources of gravity \cite{Zheng:2018fyn}, the radial component of the stress-energy tensor serves as a thermodynamic pressure: $P=T_r^r|_{r_+}$. For the EsGB theory with a quadratic coupling, the radial gravitation equation reads
\begin{equation}
16\pi T_r^r=-\left[\frac{2(1-B(r))}{r^2}+\frac{B(r)\phi'(r)^2}{2}\right]+\left[\frac{2}{r}+\frac{8\alpha(1-3B(r))\phi(r)\phi'(r)}{r^2}\right]\frac{B(r)}{A(r)}A'(r).
\end{equation}
Thus, the horizon equation of state can be written as
\begin{equation}
P=C(r_+)T_H+D(r_+)
\end{equation}
where
\begin{equation}
\begin{aligned}
C(r_+)&=\sqrt{\frac{B(r_+)}{A(r_+)}}\left[\frac{1}{2r_+}+\frac{2\alpha(1-3B(r_+))\phi(r)\phi'(r)}{r_+^2}\right],\\
D(r_+)&=-\left[\frac{1-B(r_+)}{8\pi r_+^2}+\frac{B(r_+)\phi'(r_+)^2}{32\pi}\right].
\end{aligned}
\end{equation}
Subsequently, the horizon entropy can be obtained through the following relation
\begin{equation}\label{eq:S}
S=\int V'(r_+)C(r_+)dr_+
\end{equation}
where $V(r_+)=4\pi r_+^3/3$ is the geometric volume of black holes \cite{Parikh2006}. Plugging our analytical expressions for $A(r), B(r)$ into \eqref{eq:S}, it is easy to find the integrand in \eqref{eq:S} decreases as $p$ increases, which shown in Fig.\ref{fig:7}.
This indicates that in the quadratic case, the horizon entropy of scalarized black holes is always smaller than that of Schwarzschild. Thus, it is obvious that Schwarzschild black holes are more stable, while the scalarized ones should decay to Schwarzschild through emitting scalar waves. It is worth to mention that this is in good agreement with the results in \cite{Blazquez-Salcedo:2018jnn}, which found that in the quadratic EsGB gravity, scalarized black holes are unstable under radial perturbations. In addition, like the Penrose process in general relativity, which can convert Kerr black holes into Schwarzschild black holes, it seems that in the quadratic case, a Penrose-like process may also exist to convert scalarized black holes into Schwarzschild black holes. We shall give a brief discussion in next section to show that there is a negative energy state for test particles outside the event horizon of scalarized black holes, which implies that it is possible to extract the energy of scalar charge.

\subsection{The possibility to extract the energy of scalar charge}
\begin{figure}        
	\centering 
	\includegraphics[width=0.5\textwidth]  {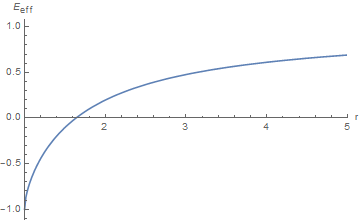}          
	\caption{The effective potential $E_{eff}$ as a function of $r$ for $\alpha=0.1, r_0=1, p=0.5$. Given the analytical approximation \eqref{eq:approx1},  $E_{eff}$ is negative near the horizon and becomes positive when $r \gtrsim 1.6$.}
	\label{fig:8}
\end{figure}

Since the formation of the scalarized black holes is an evasion of the no-hair theorems, one may say that the black holes can be described in terms of the scalar charge $D$, in addition to the mass $M$. Let us consider the motion of a test particle with mass $\mu$ in the scalarized black hole spacetime, here we assume that the particle also has a scalar charge $d$. Then, the following three conserved quantities can be derived:
\begin{equation}\label{eq:49}
p_0=-E=-\mu A(r)\dot{t}-\frac{dD}{r},
\end{equation}
\begin{equation}
p_\varphi=\mu r^2\mathrm{sin}\theta\dot{\phi},
\end{equation}
\begin{equation}\label{eq:51}
-A(r)\dot{t}^2+\frac{1}{B(r)}\dot{r}^2+r^2\dot{\theta}^2+r^2\mathrm{sin}^2\theta\dot{\phi}^2=-1.
\end{equation}
From Eqs. (\ref{eq:49})-(\ref{eq:51}), we have
\begin{equation}
-\frac{1}{A(r)}\left(\frac{p_0}{\mu}+\frac{dD}{\mu r}\right)^2+\frac{1}{B(r)}\dot{r}^2+r^2\dot{\theta}^2+\frac{p_\varphi^2}{\mu^2r^2\mathrm{sin}^2\theta}=-1.
\end{equation}
Settig $\theta=\frac{\pi}{2}$ and $k\dot{r} = 0$, we obtain the equation governing the "effective potential" of the orbits in the equatorial plane 
\begin{equation}
E_{eff}=-p_0=\frac{dD}{r}\pm\mu\sqrt{\left(\frac{p_\varphi^2}{\mu^2r^2}+1\right)A(r)}.
\end{equation}
Let $E_{\pm}$ denote the positive and negative branches of $E_{eff}$ respectively. Clearly, 
\begin{equation} 
E_+(r, D, d)=-E_-(r, D, -d).
\end{equation}
Note that there exists a negative energy state for the positive branch $E_+$ in the case of $dD<0$. For instance, setting $r_0\to1, \mu\to1, p_\varphi\to0,$ and $ d D\to-1$, for the metric function \eqref{eq:approx1}, the distribution of $E_{eff}$ for $p=0.5$ is shown in Fig.\ref{fig:8}.
It is obvious that near the event horizon, $E_{eff}<0$ while $E_{eff}>0$ in the far region. We shall call the region where $E_{eff}<0$ "effective scalar ergosphere". This suggests the existence of Penrose process to extract the energy of scalar charge.

\section{conclusions }
\label{sec:conclusion}
In this paper, by constructing approximate analytical representations for both the metric and the scalar field configurations of a spontaneously scalarized black hole solution in a EsGB theory with a quadratic coupling between the massless scalar field and the Gauss-Bonnet invariant, we have shown that the horizon entropy of a scalarized black hole is always smaller than that of a Schwarzschild black hole, which indicates that these scalarized black holes may decay to Schwarzschild black holes by emission of scalar waves. This fact also implies the possibility to extract the energy of scalar charges. 

It is also found that in a large region outside the event horizon, the analytical representations with continue-fraction expansions up to second order fit well with the numerical solution with accuracy of a fraction of 2 percent. And it is enough to compute thermodynamic quantities. Especially, from  the horizon entropy calculated from the analytically approximate solution of the metric and the scalar field configuration, we can infer  that the scalarized black holes would be thermodynamically unstable, which is in good agreement with the previous results from the entirely numerical analysis \cite{Blazquez-Salcedo:2018jnn}. These results indicate that the continue-fraction approximation is a valid approach to study the properties of numerical black hole solution.

 The dynamical stability of the scalarized black holes in the EsGB theories can also be directly investigated by employing the above approximately parameterization method. Furthermore, the authors in \cite{Gao:2018acg} indicates that in the rotating case, a regime like (10) also exists, which may finally lead to scalarization for the Kerr black holes, and such spontaneously scalarized rotating black holes have been obtained  numerically in the EsGB theory \cite{Cunha:2019dwb}. The application of the parameterized method to the rotating black hole solution deserves new work in the future.

\begin{acknowledgments}
We thank Y. Huang, A. Zhidenko, R. A. Konoplya, P. Kanti, A. Bakopoulos and H. L$\ddot{u}$ for helpful discussions. This work is supported the Program of Shanghai Normal University under grant No. KF201813.
\end{acknowledgments}

\bibliographystyle{apsrev4-1}
\bibliography{aaRefs} 

\end{document}